\documentclass[10pt,a4paper]{article}            % DO NOT DELETE THIS LINE
\usepackage{geometry}                % See geometry.pdf to learn the layout options. There are lots.
\usepackage{fancyhdr}
\usepackage{authblk}
\usepackage{graphicx}
\usepackage{url}
\makeatletter
\def\url@leostyle{%
  \@ifundefined{selectfont}{\def\UrlFont{\sf}}{\def\UrlFont{\small\ttfamily}}}
\makeatother
\usepackage{natbib}
\usepackage{epstopdf}
\usepackage{amssymb}
\usepackage[fleqn]{amsmath}
\def\mathbi#1{\textbf{\em #1}}
\numberwithin{equation}{section}
\DeclareMathSymbol{\Gamma}{\mathalpha}{letters}{"00}
\DeclareMathSymbol{\Lambda}{\mathalpha}{letters}{"03}
\DeclareMathSymbol{\Omega}{\mathalpha}{letters}{"0A}
\DeclareMathAlphabet{\mathitbf}{OML}{cmm}{b}{it}

\begin{document}                  % DO NOT DELETE THIS LINE

     % The title of the paper. Use \shorttitle to indicate an abbreviated title
     % for use in running heads (you will need to uncomment it).

\title{The Geometry of Niggli Reduction III: SAUC -- Search of Alternate Unit Cells }
%\shorttitle{Short Title}

     % Authors' names and addresses. Use \cauthor for the main (contact) author.
     % Use \author for all other authors. Use \aff for authors' affiliations.
     % Use lower-case letters in square brackets to link authors to their
     % affiliations; if there is only one affiliation address, remove the [a].
\author[1]{Keith J. McGill}
\author[1]{Mojgan  Asadi}
\author[1]{Maria Toneva Karakasheva}
\author[2]{Lawrence C. Andrews}
\author[1,*]{Herbert J. Bernstein}
\affil[1]{Dowling College, 1300 William Floyd Pkwy, Shirley, NY 11967 USA}
\affil[2]{Micro Encoder Inc., 11533 NE 118th St., Kirkland, WA 98034 USA}
\affil[*]{To whom correspondence should be addressed. Email: {\it yaya@dowling.edu}}
     % Use \shortauthor to indicate an abbreviated author list for use in
     % running heads (you will need to uncomment it).

%\shortauthor{Soape, Author and Doe}

     % Use \vita if required to give biographical details (for authors of
     % invited review papers only). Uncomment it.

%\vita{Author's biography}

     % Keywords (required for Journal of Synchrotron Radiation only)
     % Use the \keyword macro for each word or phrase, e.g. 
     % \keyword{X-ray diffraction}\keyword{muscle}

%\keyword{keyword}

     % PDB and NDB reference codes for structures referenced in the article and
     % deposited with the Protein Data Bank and Nucleic Acids Database (Acta
     % Crystallographica Section D). Repeat for each separate structure e.g.
     % \PDBref[dethiobiotin synthetase]{1byi} \NDBref[d(G$_4$CGC$_4$)]{ad0002}

%\PDBref[optional name]{refcode}
%\NDBref[optional name]{refcode}

\maketitle                        % DO NOT DELETE THIS LINE

\begin{abstract}

A crystallographic cell is a representation of a lattice, but each lattice can
be represented just as well by any of an infinite number of such unit cells.
Searching for matches to an experimentally determined crystallographic unit cell
in a large collection of previously determined unit cells is a useful
verification step in synchrotron data collection and can be a screen for ``similar'' 
structures, but it is more useful to search for a match to the lattice represented
by the experimentally determined cell.  For identification of substances with small 
cells, a unit cell match may be sufficient for unique identification.  Due to experimental error 
and multiple choices of cells and differing choices of lattice centering
representing the same lattice, simple searches based on raw cell edges and angles can 
miss similarities among lattices.  A database of lattices using the 
${\mathbi{G}^{\mathbi{6}}}$ representation of the Niggli-reduced cell as 
the search key provides a more robust and complete 
search.  Searching is implemented by finding the distance from the probe cell
to related cells using a topological embedding of the Niggli reduction in
${\mathbi{G}^{\mathbi{6}}}$, so that all cells representing similar lattices
will be found.  Comparison of results with those from older
cell-based search algorithms suggests significant value in the new approach.
\end{abstract}

     %-------------------------------------------------------------------------
     % The main body of the paper
     %-------------------------------------------------------------------------
     % Now enter the text of the document in multiple \section's, \subsection's
     % and \subsubsection's as required.

\section{Introduction}

\cite{Andrews2012} introduced a topological embedding of the Niggli ``cone'' of reduced
cells with the goal of calculating a meaningful distance between unit cells. 
In the second paper of this series, the embedding was used to determine 
likely Bravais lattices for a unit cell. 
Here we apply the embedding to searching within a database for lattices ``close'' 
to the lattice of a given probe cell.

A crystallographic cell is a representation of a lattice, but each lattice can
be represented just as well by any of an infinite number of such unit cells.
Searching for matches to an experimentally determined crystallographic unit cell
in a large collection of previously determined unit cells is a useful
verification step in synchrotron data collection and can be a screen for ``similar'' 
structures \cite{Ramraj2011} \cite{mighell2002lattice}, but it is more useful to 
search for a match to the lattice represented by the experimentally determined cell,
which may involve many more cells.  For identification of substances with small cells, a 
unit cell match may be sufficient for unique identification \cite{Mighell2001}.

Due to experimental error 
and multiple cells representing the same lattice and differing choices of lattice centering, simple searches based on raw cell 
edges and angles can miss similarities.  A database of lattices using the ${\mathbi{G}^{\mathbi{6}}}$ representation
of the Niggli-reduced cell as the search key provides a more robust and complete 
search.  Searching is implemented by finding the distances from the probe cell
to related cells using a topological embedding of the cone of Niggli reduced cells in
${\mathbi{G}^{\mathbi{6}}}$.  Comparison of results to those from older
cell-based search algorithms suggests significant value in the new approach.

\section{History}

Tabulations of data for the identification of minerals dates to the 18th and 19th centuries. Data collected included interfacial angles of crystals (clearly related to unit cell parameters) and optical effects.  See the historical review in \cite{burchard1998history}. With the discovery of x-ray diffraction, those tables were  supplanted by new collections. Early compilations that included unit cell parameters arranged for material 
identification were "Crystal Structures"  \cite{wyckoff1931structure}, "Crystal Data Determinative Tables" \cite{Donnay1943},
and “Handbook for Metals and Alloys” \cite{Pearson1958}. Early computerized searches were created by JCPDS in the mid-1960's \cite{GGJohnson2013} and the Cambridge Structural Data file and its search programs \cite{Allen1973}.

Those first searches were sensitive to the issues of differing equivalent presentations
of the same lattice. The first effective algorithm for resolving that issue
was \cite{Andrews1980} using the V7 algorithm \cite{NIH/EPA1980}.  Subsequently, other programs using
the V7 algorithm have been described (see Table 1).
The V7 algorithm has the advantage over simple Niggli-reduction based cell searches
of being stable under experimental error.  However, sensitivity to a change in an angle
is reduced as that angle nears 90 degrees.

\begin{table}
\caption{Programs designed to perform effective searches in a unit cell database}
\begin{center}
\begin{tabular}{lll}
{\bf Program}&{\bf Reference}&{\bf Method}\\ 
Cryst&\cite{Andrews1980}&V7\\
&\cite{NIH/EPA1980}&\\
cdsearch&\cite{Toby1994}&V7\\
Quest&\cite{Allen1973}&Reduced cell\\
Nearest-Cell&\cite{Ramraj2011}&Reduced cell\\
WebCSD, Conquest&\cite{Thomas2010}&${\mathbi{G}^{\mathbi{6}}}$ iterative\\
SAUC&(this work)&${\mathbi{G}^{\mathbi{6}}}$, Niggli embedding\\
\end{tabular}
\end{center}
\label{tab:programs}
\end{table}%

\section{Background}

An effective search method must find ways to search for related unit cells, 
even when they appear to be tabulated in ways that make them seem different. A trivial example is:

\[ a=10.0, b=10.01, c=20,\]  
\[ \alpha=65, \beta=75, \gamma=90\]

versus
\[ a=10.0, b=10.05, c=20,\]  
\[ \alpha=75, \beta=65, \gamma=90.\]

\noindent{}Clearly, these unit cells are almost identical, but simple tabulations 
might separate them. A somewhat more complex example includes the following primitive cells:

\[ a=3.1457, b=3.1457, c=3.1541,\]  
\[\alpha=60.089, \beta=60.0887, \gamma=60.104\]

versus
\[ a=3.1456, b=3.1458, c=3.1541,\]  
\[\alpha=90.089, \beta=119.907, \gamma=119.898\].

Here the relationship is not as obvious. The embedding of \cite{Andrews2012} 
can be used to show that the distance between these two cells is quite small 
in ${\mathbi{G}^{\mathbi{6}}}$ (0.004 {\AA}ngstrom units squared in ${\mathbi{G}^{\mathbi{6}}}$).

\section{Implementation: 1 -- Distance}

The program SAUC is structured to allow use of several alternative metrics 
for searching among cells in an attempt to identify cells representing similar
lattices.  To simplify comparisons among results with the 
different metrics, all have been linearized and normalized, 
{\it i.e.} converted to {\AA}ngstrom units and scaled to be commensurate with the $L_2$ norm given below: 

\begin{itemize}

\item{A simple $L_1$ or $L_2$ norm based on
\[ [ a, b, c, \alpha  (b + c)/2, \beta  (a + c)/2, \gamma  (a + b)/2 ] \]

with the distance scaled by $1/\sqrt{6}$ in the case of the $L_1$ norm and unscaled in case of
the $L_2$ norm.  The angles are assumed to be in radians and the edges in {\AA}ngstroms.
The angles were converted to {\AA}ngstroms by multiplying by the average of the relevant
edge lengths.
}
\item{The square root of the BGAOL Niggli cone embedding distance NCDist based on
\[ [ a^2, b^2, c^2, 2 b c cos(\alpha) , 2 a c cos(\beta), 2 a b cos (\gamma) ] \]
with the distances scaled by  $1/\sqrt{6}$ and divided by the reciprocal of the
average length of cell edges f.  The square root linearizes the
metric to {\AA}ngstrom units}.

\item{The V7 distances based on individual components linearized to {\AA}ngstrom units
\[ [ a, b, c, 1/a^{*}, 1/b^{*}, 1/c^{*}, V^{1/3} ] \]
and scaled by $\sqrt{6/7}$.  $V$ is the volume.}

\end{itemize}

These metrics are applied to reduced primitive cells $[ a, b, c, \alpha, \beta, \gamma ]$ and,
when the reciprocal cell $[ a^{*}, b^{*}, c^{*}, \alpha^{*}, \beta^{*}, \gamma^{*} ]$ is
needed for the V7 metric, that cell is also reduced.

In order to facilitate comparisons to older searches that just consider simple
ranges in  $[ a, b, c, \alpha, \beta, \gamma ]$, an option for such searches was also
included in SAUC.

\subsection{Validity of using the square root}

The use of the square root on a metric preserves the triangle inequality, which is
important in order to preserve the metric as a metric-space ``metric''.  The triangle
inequality states that for any triangle, the sum of the lengths of any two sides
is greater than the length of the third side.   In metric space terms, the metric
$d(x,y)$ of a metric space $M$ satisfies $d(x,z) \leqslant d(x,y) + d(y,z),
\text{~~~}\forall x,y,z \in M$.  Suppose a function $f$ satisfies the following conditions:
\[
u \geqslant v \Rightarrow f(u) \geqslant f(v),\text{~~~} \forall u,v
\]
\[
f(u+v) \leqslant f(u)+f(v), \text{~~~}\forall u,v
\]

\noindent{} then, if d(x,y) satisfies the triangle inequality, f(d(x,y)) will also
satisfy the triangle inequality:
\[
d(x,z) \leqslant d(x,y) + d(y,z) \]
\[ \Rightarrow f(d(x,z))  \leqslant f(d(x,y) + d(y,z)) \leqslant f(d(x,y)) + f(d(y,z))
\]

The square root satisfies the stated requirements.  It is monotone, and
\[
\sqrt{u+v} \leqslant \sqrt{u}+\sqrt{v}
\]
\[
\Leftrightarrow u+v \leqslant (\sqrt{u}+\sqrt{v})^2 = u+v+2\sqrt{uv}
\]
which is clearly true.

\section{Implementation: 2 -- Searching}

Range searching in a mapped embedding needs to be done using a nearest-neighbor algorithm (or ``post-office problem'' algorithm \cite{Knuth1973}).
Exact matches are unlikely since most unit cells representing lattices in a database are experimental, 
and probe cells are also likely have been calculated from experimental data. Several efficient 
algorithms are available; we have used an implementation of neartree \cite{Andrews2001}.

The raw unit cell data is loaded into the tree once and serialized to a dump file on disk; subsequent
searches do not need to wait for the $O(N log(N))$ tree build, which for the $70,000+$ cells from the
PDB can take half an hour in the BGAOL NCDist metric.  The linearization makes the search space more
compact and reduces the tree depth, thereby speeding searches.  Because the PDB unit cell database contains
many identical cells, we modified NearTree to handle the duplicates in auxiliary lists,
further reducing the tree depth and speeding searches.

\section{Comparison of Search Methods}

The simplest approach to lattice searching is a simple box search 
on ranges in unit cell $a$, $b$ and $c$ and
possibly on $\alpha$, $\beta$ and $\gamma$, as for example 
in the ``cell dimensions'' option in
the RCSB advanced search at \url{http://www.rcsb.org/pdb/search/advSearch.do}) for the Protein Data Bank \cite{Berman2000}.  In the following examples, we will call that type of search ``Range''.   For the reasons
discussed above, such simple searches can fail to find unit cells with very different angles
that actually represent similar lattices.   Such searches are best characterized as
cell searches, rather than as lattice searches.

Searching on primitive reduced cells greatly improves
the reliability of a search, as for example in \cite{Ramraj2011} 
at \url{http://www.strubi.ox.ac.uk/nearest-cell/nearest-cell.cgi}, which uses a metric based on the reduced cell
and all permutations of axes.   While an improvement over simple range searches as 
discussed above, such searches can also miss similar lattices if the number of 
alternate lattice presentations considered is not complete.  One way to reduce such gaps in searches is to use only parameters that do not depend on
the choice of reduced presentation.  The \cite{Andrews1980} approach using 7 parameters (three
reduced cell edges, three reduced reciprocal cell edges and the volume), ``V7'', helps, but has
difficulty distinguishing cells with angles near 90 degrees.  The NCDist approach used here, derived from
\cite{Andrews2012}, both fills in the gaps and handles angles near to 90 degrees. 

Consider, for example, the unit cells of phospholipase $A_2$ discussed by \cite{Trong2007}.
They present three
alternate cells from three different PDB entries that are actually for the same structure: 
 
\noindent{}$[57.98,57.98,57.98,92.02,92.02,92.02]$ from entry 1FE5 \cite{1fe5} in space group $R32$, 

\noindent{}$[80.36,80.36,99.44,90,90,120]$ from
entry 1U4J \cite{1u4j} in space group $R3$ and 

\noindent{}$[80.949,80.572,57.098,90.0,90.35,90.0]$ from entry 1G2X \cite{1g2x} in space group
$C2$.  No simple range search can bring these three cells together.   For example, if we
use the PDB advanced cell dimensions search around the cell from IU4J with 
edge ranges of $\pm 3$ {\AA}ngstroms and
angle ranges of $\pm 1$ degree, we get 28 hits: 1CG5, 1CNV, 1FW2, 1G0Z, 1GS7, 1GS8, 1HAU, 1ILD, 1ILZ, 1IM0, 1LR0, 1NDT, 1OE1, 1OE2, 1OE3, 
1QD5, 1U4J, 2BM3, 2BO0, 2H8A, 2HZ5, 2OHG, 2REW, 2WCE, 3I06, 3KKU, 3Q98, 3RP2, of which only three actually have cells close to the target using
the linearized NCDist metric :  2WCE at 2.96 {\AA}ngstroms, 1G0Z at 0 {\AA}ngstroms,
and 1U4J, the target itself.  The remaining cells are, as we will see, rejected
under the Nearest-Cell and the V7 metric.  The simple Range searches are not appropriate
to this problem.

Table \ref{tab:Stenkamp} shows
partial results from a lattice search using Nearest-Cell, and a V7 
search using SAUC and a NCDist search using SAUC.  We have restricted the 
searches to NCDist distances $\leqslant 3.5$ {\AA}ngstroms.  The Nearest-Cell metric appears to be in $\AA^2$. 
The column with the square root of the Nearest-Cell metric
facilitates comparison with the linearized SAUC V7 and NCDist metrics.  
The searches showed consistent behavior:  The three cells noted by 
\cite{Trong2007} are found in the same relative positions  by all three 
searches.  All cells found by Nearest-Cell are also found by both V7 and 
NCDist.  Of the 42 structures found by all three metrics within 
3.5 {\AA}ngstroms under the NCDist metric, four (1G0Z, 1G2X, 1DPY and 1FE5) 
are E.C class 3.1.1.4 phospholipase A2 structures, and three (1PKR, 1SGC and
1VRI) are other hydrolases (E. C. classes 3.4.21.7, 3.4.21.80, and 3.4.19.2,
respectively)  However, ten cells found by V7 and NCDist 
were not found by  Nearest-Cell (2OSN, 2CMP, 3MIJ, 2SGA, 2YZU, 3SGA, 4SGA,
5SGA, 1CDC and 2CVK).  Of those ten, one (2OSN) is an E.C class 3.1.1.4 
phospholipase A2 structure and four (2SGA, 3SGA, 4SGA and 5SGA are hydrolases,
specifically E.C. class 3.4.21.80 proteinase A.  Two of the ten (2YZU and
2CVK) are thioredoxin, for which the ProMOL \cite{Craigsubmitted} motif finder shows significant
active site homologies to multiple hydrolase motifs (2YZU has site homologies to 132L, 135L and 
1LZ1 in E.C. class 3.2.1.17 and to 4HOH in E.C. class 3.1.27.3, 2CVK to 
1AMY in E.C. class 3.2.1.1, to 1BF2 in E.C. class 3.2.1.68, to 1EYI in
class 3.2.3.11, {\it etc.}).  For 1CDC, a ``metastable structure of CD2'',
proMOL shows an active site homology to 1ALK of E.C. class 3.1.3.1, another
hydrolase.

\begin{table}
\caption{Comparison of search results for cell $[80.36,80.36,99.44,90,90,120]$ from 
entry 1U4J in space group $R3$.  See \cite{Trong2007}.   In each case the PDB ID \cite{Bernstein1977} \cite{Berman2000} found is shown with the
distance metric for that method. In the case of Nearest-Cell \cite{Ramraj2011}, a second column with
the square root of the metric is provided as well.   The results are sorted by the NCDist distance.  Results have
been cut off at 3.5 {\AA}ngstroms in the NCDist metric.  The three alternate cells cited by \cite{Trong2007} are marked with ``(*)''.
The Nearest Cell results are from the {http://www.strubi.ox.ac.uk/nearest-cell/nearest-cell.cgi} web site. 
The NCDist and V7 results are from SAUC.}
\begin{center}
{\footnotesize \begin{tabular}{lllllll}
            &{\bf Nearest-}&{\bf Sqrt of}&&&&\\
            &{\bf Cell}&{\bf Nearest-}&{\bf V7}&{\bf NCDist}&&\\
{\bf PDB ID}&{\bf metric}&{\bf Cell}&{\bf metric}&{\bf metric}& {\bf Molecule}&{\bf E.C. Code}\\
{\bf 1U4J (*)}&0&0&0&0&Phospholipase A2 isoform 2&3.1.1.4\\
1G0Z&0&0&0&0&Phospholipase A2&3.1.1.4\\
{\bf 1G2X (*)}&0.11&0.33&0.2&0.9&Phospholipase A2& 3.1.1.4\\
2OSN&&&0.2&0.9&Phospholipase A2 isoform 3&3.1.1.4\\
2CMP&&&0.7&1.5&Terminase small subunit&\\
3KP8&0.43&0.66&1.1&1.7&VKORC1/Thioredoxin domain protein&\\
3MIJ&&&1&1.7&RNA (5'-R(*UP*AP*GP*GP*GP*UP&\\
    &&& &   &~~~~~~~~~~*UP*AP*GP*GP*GP*U)-3')&\\
3E56&0.4&0.63&1.5&1.9&Putative uncharacterized protein\\
1CSQ&0.49&0.7&1.8&2&Cold Shock Protein B (CSPB)\\
3SVI&0.54&0.73&1.9&2.1&Type III effector HopAB2\\
1FKF&0.83&0.91&2.7&2.4&FK506 binding protein&5.2.1.8\\
1FKJ&0.83&0.91&2.7&2.4&FK506 binding protein&5.2.1.8\\
1BKF&0.91&0.95&2.8&2.5&Subtilisin Carlsberg&3.4.21.62\\
1FKD&0.86&0.93&2.8&2.5&FK506 binding protein&5.2.1.8\\
2FKE&0.91&0.95&2.9&2.6&FK506 binding protein&5.2.1.8\\
3TJY&0.88&0.94&3&2.6&Effector protein hopAB3&\\
2I5L&1.06&1.03&3.7&2.7&Cold shock protein cspB&\\
2WCE&1.21&1.1&3.5&3&Protein S100-A12&\\
3P63&1.28&1.13&4&3&Ferredoxin&\\
1F9P&1.37&1.17&4.7&3.1&Connective tissue activating peptide-III&\\
2CXD&1.36&1.17&4.7&3.1&Conserved hypothetical&\\
    &    &    &   &   &~~~~~~~~protein, TTHA0068&\\
2SGA&&&4.9&3.1&Proteinase A&3.4.21.80\\
2YZU&&&4.8&3.1&Thioredoxin&\\
3SGA&&&5&3.1&Proteinase A (SGPA)&3.4.21.80\\
4SGA&&&4.8&3.1&Proteinase A (SGPA)&3.4.21.80\\
5SGA&&&4.9&3.1&Proteinase A (SGPA)&3.4.21.80\\
1GUS&1.15&1.07&0.3&3.2&Molybdate binding protein II\\
1PKR&1.44&1.2&4.7&3.2&Plasminogen&3.4.21.7\\
1SGC&2.45&1.57&5.2&3.2&Proteinase A&3.4.21.80\\
2VRI&1.5&1.22&5.2&3.2&Non-structural protein 3&3.4.19.12\\
1CDC&&&4.9&3.3&CD2&\\
1DPY&1.24&1.11&2.3&3.3&Phospholipase A2&3.1.1.4\\
{\bf 1FE5 (*)}&1.24&1.11&2.3&3.3&Phospholipase A2&3.1.1.4\\
1GUT&1.2&1.1&0.1&3.37&Molybdate binding protein II&\\
2C9Q&1.46&1.21&4.8&3.3&Copper resistance protein C&\\
2CVK&&&5.2&3.3&Thioredoxin&\\
2HE2&1.48&1.22&4.8&3.3&Discs large homolog 2&\\
2IT5&1.62&1.27&5.6&3.3&CD209 antigen, DCSIGN-CRD&\\
3SU1&1.59&1.26&5.2&3.3&Genome polyprotein&\\
3SU5&1.58&1.26&5.1&3.3&NS3 protease, NS4A protein&\\
3SU6&1.52&1.23&5&3.3&NS3 protease, NS4A protein&\\
1SL4&1.68&1.3&5.8&3.4&mDC-SIGN1B type I isoform&\\
2IT6&1.73&1.32&6&3.4&CD209 antigen&\\
3CYO&1.81&1.35&5.6&3.4&Transmembrane protein&\\
3SU2&1.6&1.26&5.2&3.4&Genome polyprotein&\\
3SU3&1.64&1.28&5.3&3.4&NS3 protease, NS4A protein&\\
1H9M&1.18&1.09&1.1&3.5&Molybdenum-binding-protein&\\
1X90&1.34&1.16&5.2&3.5&Invertase/pectin methylesterase&\\
    &    &    &   &   &~~~~~~~~~~~~~inhibitor family protein&\\
2E6L&1.78&1.33&5.8&3.5&Nitric oxide synthase, inducible&1.14.13.39\\
3CP1&1.98&1.41&6.1&3.5&Transmembrane protein&\\
3SU0&1.75&1.32&5.7&3.5&Genome polyprotein&\\
3SV6&1.74&1.32&5.6&3.5&NS3 protease, NS4A protein&\\
3SV7&1.73&1.32&5.6&3.5&NS3 protease, NS4A protein&\\
\end{tabular}}
\end{center}
\label{tab:Stenkamp}
\end{table}%

The significant gaps in the Nearest-Cell search do not appear to be a problem of 
the distance for the Nearest-Cell 
search having been cut off at a too-small value.  For the common hits between 
the square root of the Nearest-Cell metric and the linearized NCDist metric, 
a linear fit is excellent, with $R^2 = 0.89$
and no points are very far from the line.   The agreement of the linearized V7 to the other two
metrics is much noisier because of loss of sensitivity of the V7 metric for angles
near 90 degrees and the inherent difficulty the V7 metric has in discriminating between
the $+++$ and $---$ parts of the Niggli cone.  For example, 1GUT \cite{igut} is at distances 1.2
and 3.7 from 1UJ4 in the Nearest-Cell and linearized NCDist metrics, respectively, but 
only 0.1 in the V7 metric.  The 1GUT cell is 

\noindent{}$[78.961,82.328,57.031,90.00,93.44,90.00]$ in C 1 2 1, Z=24,

\noindent{}with a primitive cell 

\noindent{}$[57.031, 57.0367, 57.0367, 92.3918, 92.3804, 92.3804]$

\noindent{}which corresponds to a ${\mathbi{G}^{\mathbi{6}}}$ vector 

\noindent{}$[3252.53, 3253.18, 3253.18, -271.53, -270.208, -270.208]$

\noindent{}and a linearized V7 vector 

\noindent{}$[52.8004, 52.8057, 52.8057, 52.7101, 52.7101, 52.7053, 52.7569].$

\noindent{}The 1U4J cell is 

\noindent{}$[80.36, 80.36,99.44, 90, 90, 120]$ in R3, Z=18,

\noindent{}with a primitive cell

\noindent{}$[57.02, 57.02, 57.02, 89.605, 89.605, 89.605]$

\noindent{}which corresponds to a ${\mathbi{G}^{\mathbi{6}}}$ vector 

\noindent{}$[3251.28, 3251.28, 3251.28, 44.8265, 44.8265, 44.8265]$

\noindent{}and a linearized V7 vector 

\noindent{}$[52.7902, 52.7902, 52.7902, 52.7878, 52.7878, 52.7878, 52.789]$

\noindent{}This is almost identical to the 1GUT V7 vector, even though the corresponding primitive cells 
and ${\mathbi{G}^{\mathbi{6}}}$ cells differ significantly. 

\section{SAUC program availability}

SAUC is an open source program released under the GPL and LGPL on Sourceforge
in the iterate project at

\noindent{}\url{http://sf.net/projects/iterate/}

A recent release is available at

\noindent{}\url{http://downloads.sf.net/iterate/sauc-0.6.tar.gz}

A web site, shown in Figs. \ref{fig:website} and \ref{fig:webresult}, and on which searches may be done and from which the latest
release may be retrieved is available at

\noindent{}\url{http://www.bernstein-plus-sons.com/software/sauc}

\begin{figure} 
\caption{Query box from a SAUC website at {http://www.bernstein-plus-sons.com/software/sauc/} } 
\scalebox{.45}{\includegraphics{website.epsf}} 
\label{fig:website}
\end{figure} 

\begin{figure} 
\caption{Partial results from SAUC website query} 
\scalebox{.39}{\includegraphics{webresult.epsf}} 
\label{fig:webresult}
\end{figure}

     %-------------------------------------------------------------------------
     % The back matter of the paper - acknowledgements and references
     %-------------------------------------------------------------------------

     % Acknowledgements come after the appendices

\section*{Acknowledgements}

The authors acknowledge the invaluable assistance of Frances C. Bernstein.
\\
The work by Herbert J. Bernstein, Keith J. McGill, 
Mojgan Asadi and Maria Toneva Karakasheva has been supported 
in part by NIH NIGMS grant GM078077.
The content is solely the responsibility of the authors and does not
necessarily represent the official views of the funding agency.
\\
Lawrence C. Andrews would like to thank Frances and Herbert Bernstein for hosting him during 
hurricane Sandy and its aftermath. Elizabeth Kincaid has contributed significant support in many ways.
\\
Our thanks to Ronald E. Stenkamp for pointing us to the highly relevant work in \cite{Trong2007}.
 
     %-------------------------------------------------------------------------
     % TABLES AND FIGURES SHOULD BE INSERTED AFTER THE MAIN BODY OF THE TEXT
     %-------------------------------------------------------------------------

\onecolumn

     % Postscript figures can be included with multiple figure blocks

\bibliographystyle{abbrvnat}
\bibliography{BGAOL_refs_23May13}

\end{document}